\documentclass[aps,prl,twocolumn,10pt,superscriptaddress,nofootinbib,nobibnotes,longbibliography]{revtex4-1}

\usepackage{amssymb}
\usepackage{graphicx}
\usepackage{amsmath}
\usepackage[breaklinks=true]{hyperref}
\usepackage{xurl}
\usepackage{subfigure}
\usepackage{multirow}
\usepackage{setspace}
\usepackage{verbatim}
\usepackage{float}
\usepackage{color}
\usepackage[normalem]{ulem}
\usepackage[utf8]{inputenc}
\usepackage[table,xcdraw]{xcolor}
\usepackage{makecell}
\usepackage{url}
\usepackage{bm}

\begin{document}

\title{Quantum-information bounds on black-hole evaporation and de Sitter expansion}


\author{Rong-Gen Cai}
\email{caironggen@nbu.edu.cn}
\affiliation{Institute of Fundamental Physics and Quantum Technology, \& School of Physical Science and Technology, Ningbo University, Ningbo, 315211, China}

\author{Li Hu}
\email{huli@nbu.edu.cn}
\affiliation{Institute of Fundamental Physics and Quantum Technology, \& School of Physical Science and Technology, Ningbo University, Ningbo, 315211, China}

\author{Shao-Jiang Wang}
\email[Corresponding author:]{schwang@itp.ac.cn}
\affiliation{Institute of Theoretical Physics, Chinese Academy of Sciences (CAS), Beijing 100190, China}
\affiliation{Asia Pacific Center for Theoretical Physics (APCTP), Pohang 37673, Korea}

\begin{abstract}
It is well-known that the unitary nature of black hole evaporation enforces its entropy evolution to follow the Page curve. In this Letter, we further explore the outcome of the quantum speed limit on how fast a quantum system can evolve unitarily. When combined with the maximal information transmission bound, we place upper limits on the evaporation rates of dynamical black-hole entropy and mass that are otherwise divergent at the very end of the Hawking evaporation, during which the Penrose inequality from the cosmic censorship conjecture is consistently obeyed. Further applying the maximal information transmission bound to an effective-field-theory description for all sub-horizon regions of a de Sitter space would sufficiently reproduce a weaker trans-Planckian censorship conjecture up to a numeric prefactor conditionally.
\end{abstract}
\maketitle

\textit{\textbf{Introduction.}---} 
What is a black hole (BH)? This might be the most inspiring question in the history of gravitation. In classical Newtonian gravity, the existence of the ``BH'' was essentially predicted by Laplace in 1796~\cite{BH1796} from the escape velocity $v_\mathrm{max}=\sqrt{2GM/R}$ of a particle to be static at infinity~\footnote{Strictly speaking, what Laplace found is actually a ``gray'' hole as particles can reach a non-vanishing velocity at a finite distance.} in a potential well of a non-luminous body of size $R$ and mass $M$, that is, requiring this escape velocity to be no faster than the speed of light leads to a minimal size $R\geq2GM\equiv R_s$, which was recognized more than a hundred years later as the Schwarzschild radius~\cite{Schwarzschild:1916uq,Schwarzschild:1916ae}. Although the derivation above is totally wrong from a modern point of view, what essentially matters is that the BH has the largest compactness $C\equiv M/R\leq M/R_s=1/2G$. In classical Einstein gravity, the concept of escape velocity reaching light speed has been replaced by the maximal trapped null surface, and hence the singularity theorem~\cite{Penrose:1964wq} of gravitational collapse has confirmed that~\cite{Cardoso:2019rvt}, in classical gravity, \textit{a BH is the most compact object in nature}.

In semi-classical gravity, the analog between the BH dynamics and thermodynamics has inspired Bekenstein~\cite{Bekenstein:1972tm,Bekenstein:1973ur} and Hawking~\cite{Bekenstein:1973ur,Hawking:1975vcx} to endow a BH with entropy $S_\mathrm{BH}=A/4G$ and temperature $T_H=\kappa/2\pi$ from its horizon area $A=4\pi R_s^2$ and surface gravity $\kappa=1/4GM$, respectively. For a weakly gravitating system of internal energy $E$ and maximal size $R$, Bekenstein~\cite{Bekenstein:1980jp} has conjectured a more general but heuristic entropy bound $S_B\leq2\pi RE$ that was later specified by Casini~\cite{Casini:2008cr,Blanco:2013joa} as a result of the positivity of relative entropy. Together with the maximal compactness of BH, the saturation of the Bekenstein entropy bound seems to be equivalent to the holographic bound~\cite{tHooft:1993dmi,Susskind:1994vu}, inferring the largest surface entropy density $S_B/4\pi R^2\leq(1/2)M/R\leq1/4G$, that is, \textit{a BH is the densest hard disk in nature}.  

In quantum gravity yet to be uncovered, a BH usually serves as a gravitational dual descriptions~\cite{tHooft:1993dmi,Susskind:1994vu,Maldacena:1997re,Witten:1998qj,Gubser:1998bc} for field theories in non-perturbative, non-equilibrium, or non-linear regimes~\cite{Hubeny:2010ry,Liu:2018crr}. 
For strongly coupled field theories with a gravitational dual description involving a BH in anti-de Sitter (AdS) space, the ratio of shear viscosity to entropy density saturates the Kovtun-Son-Starinets (KSS) bound~\cite{Kovtun:2004de} $\eta/s\geq1/4\pi$, suggesting that \textit{a BH is the most ideal fluid in nature}. 
For a perturbed BH relaxing back to an equilibrium state, the Bekenstein-Hod bound~\cite{Hod:2006jw} on dissipation time $t_d\geq1/\pi T$ is saturated from LIGO-Virgo measurements~\cite{Carullo:2021yxh} on the imaginary part $\omega_I\sim1/t_d$ of fundamental quasi-normal modes during the ringdown phase of BH binary merger, suggesting that \textit{a BH is the fastest dissipater in nature}. 
For a quantum many-body system, the thermal-averaged out-of-time-order correlator (OTOC) $C(t)=-\langle[W(t),V(0)]^2\rangle$ of local Hermitian operators separated in time can grow exponentially in time $C(t)\sim\exp(2\lambda_Lt)$ as a quantum analogue of classical chaos~\cite{Shenker:2013pqa}, where a holographic bound~\cite{Maldacena:2015waa} on the quantum Lyapunov exponent $\lambda_L\leq2\pi T$ is saturated for an AdS BH in gravitational dual description of a strongly-coupled system with large degrees of freedom. Hence, \textit{a BH is the most chaotic system in nature}.

In quantum information that may be intimately related to quantum gravity~\cite{Hayden:2007cs}, the spread of quantum information can be diagnosed by measuring the scrambling time $t_*$ with $C(t_*)\sim\mathcal{O}(1)$~\cite{Swingle:2016var} characterizing how fast the interactions cause initially commuting operators $C(t)$ to fail to commute, which is similar to but totally different from (also significantly longer than) the dissipation time $t_d$ from the exponential decay time of two-point correlator $\langle V(0)V(t)\rangle$. The shortest scrambling time $t_*\sim t_d\log S$ is logarithmic in entropy, with such a saturation realized for a BH~\cite{Sekino:2008he,Lashkari:2011yi}. Therefore, \textit{a BH is the fastest scrambler in nature}. 

In quantum mechanics, the modern interpretation for the Heisenberg's uncertainty relation of time and energy sets a fundamental bound on how fast a quantum system can evolve unitarily, which is characterized by the so-called quantum speed limit (QSL) (see, e.g., Ref.\cite{Deffner:2017cxz} for a historical review and references therein) on the minimal time a quantum system needs to evolve between two distinguishable states, $\delta t\geq t_\mathrm{QSL}\equiv\pi\hbar/2\langle E\rangle$, in terms of the average of Hamiltonian energy $\langle E\rangle$~\cite{Margolus:1997ih}. For a relatively small spread in energy $\delta E$, this QSL should be taken as $t_\mathrm{QSL}\equiv(\pi\hbar/2)\max(1/\delta E,1/\langle E\rangle)$. 
In quantum computation, this QSL also sets the maximal rate with which quantum information can be processed, that is, to perform an elementary logical operation in time $\delta t$ requires an average amount of energy $\langle E\rangle\geq\pi\hbar/2\delta t$, leading to the Lloyd’s bound~\cite{Lloyd:2000cry} on the computational complexity $\mathrm{d}\mathcal{C}/\mathrm{d}t\leq2\langle E\rangle/\pi\hbar$, i.e., a system with an average energy $\langle E\rangle$ can perform a maximum of $2\langle E\rangle/\pi\hbar$ logical operations per second, which is saturated for a Schwarzschild BH~\cite{Brown:2015bva,Brown:2015lvg}. Therefore, \textit{a BH is the fastest computer in nature.} 
In quantum communication, this QSL also sets a maximal rate at which information (entropy) can be transferred, $|\mathrm{d}S/\mathrm{d}t|\leq\pi k_B E/\hbar$~\cite{Deffner:2019ivt}, which was initially proposed by Bremermann~\cite{Bremermann:1967} and refined by Bekenstein~\cite{Bekenstein:1981zz} (but still with an ambiguity for the specification of message energy $E$ as we will clarify later in this Letter) from the Bekenstein entropy bound~\cite{Bekenstein:1980jp} plus special relativity.

This raises a natural but profoundly deep question, as before, whether this Bremermann-Bekenstein bound on information transfer can be saturated by a BH. It turns out as a nice surprise that~\cite{Wang:2023wsm} naively applying this Bremermann-Bekenstein bound on the entropy changing rate to the would-be divergent decreasing rate of Bekenstein-Hawking entropy near the end of BH evaporation would necessarily conjectured a constant-rate upper-bound for mass evaporation, which is consistent with the Penrose inequality~\cite{1973NYASA.224..125P,Ben-Dov:2004lmn,Mars:2009cj}, a manifestation of the cosmic censorship conjecture~\cite{Penrose:1969pc,Hawking:1970zqf,Penrose:1999vj}. With a large number of all possible kinds of Hawking radiation particles to be emitted, such a conjectured transition period from an accelerating evaporation to a constant evaporation in BH mass could even occur in the semi-classical regime before quantum gravity could come into play. 

In this letter, we will rigorize this observation for a dynamically evaporating BH~\cite{Hollands:2024vbe} and further generalize it to the earlier stage of Hawking evaporation where the dynamical BH entropy is conjectured to decrease at a rate analogue to the Pendry's bound up to a numeric prefactor~\cite{Pendry1983}, which is just the Bremermann-Bekenstein bound when the message energy is much smaller than the average energy of the system. Therefore, we propose a time-derivative curve for the upper bound on the entropy evolution due to the unitary evolution bounded by the QSL. As a byproduct, we also apply the Bremermann-Bekenstein bound to all sub-horizon regions of an expanding de Sitter (dS) space with an effective-field-theory description, which is consistent with the trans-Planckian censorship conjecture up to a numeric prefactor.

\textit{\textbf{Dynamical evaporation}.---}
The previous study~\cite{Wang:2023wsm} relies on the quasi-static assumption during evaporation, with the usual BH thermodynamics at equilibrium forced to apply. This can be relaxed by considering a dynamical BH entropy as long as the physical-process first law of BH thermodynamics, $(\kappa/2\pi)\delta S=\delta M$, is respected. Here, we adopt a recent proposal~\cite{Hollands:2024vbe} of dynamical BH entropy $S_\mathrm{dyn}=A_\mathrm{app}/4G$ from the area $A_\mathrm{app}$ of the apparent horizon to first-order perturbations around a stationary BH in general relativity.
With this definition, the physical-process first law is explicitly satisfied: 
\begin{align}\label{eq:PPFL}
\Delta S_\mathrm{dyn}=8\pi GM_\mathrm{BH}\Delta M_\mathrm{BH}.
\end{align}
We rederive this relation below following general treatments~\cite{Hollands:2024vbe} but for four-dimensional Einstein gravity.

Recall that in four-dimensional Einstein gravity, the horizon $\mathcal{H}$ is a null hypersurface generated by a null vector field $k^a=\big(\frac{\partial}{\partial\lambda}\big)^a$ with $\lambda$ being an affine parameter. Meanwhile, there also exists another null basis $l^a$ for the separation to $\mathcal{H}$. The cross section $\mathcal{C}$ is a two-dimensional space-like surface with constant $\lambda$. The metric can be decomposed onto the horizon as $g_{ab}=-k_al_b-l_ak_b+q_{ab}$, where $q_{ab}$ is the induced spatial metric on the cross section. In the case of general relativity, the BH entropy $S$ that satisfies the Bekenstein-Hawking formula on an arbitrary cross section $\mathcal{C}$ can be expressed as
\begin{align}
S_\mathrm{BH}[\mathcal{C}]=\frac{1}{4G}\int_{\mathcal{C}}\bm{\epsilon},
\end{align}
where $\bm{\epsilon}=\sqrt{q}\,\bm{e}$ is the volume form, and $q$ is the trace of the induced spatial metric. After that, let $\Delta S_\mathrm{BH}$ denote the entropy difference between the two cross sections,
\begin{align}
\Delta S_\mathrm{BH}=\frac{1}{4G}\int_{\mathcal{C}}\bm{e}\int_{\lambda_1}^{\lambda_2}\frac{\mathrm{d}}{\mathrm{d}\lambda}\sqrt{q}\,\mathrm{d}\lambda.
\end{align}
Substituting the expansion of the horizon $\theta=\nabla_ak^a=\frac{1}{\sqrt{q}}\frac{\mathrm{d}}{\mathrm{d}\lambda}\sqrt{q}$ into the above equation and integrating by parts, the entropy difference can be further expressed as
\begin{align}\label{eq:IBP}
\Delta S_\mathrm{BH}
=\frac{1}{4G}\int_{\mathcal{C}}\bm{\epsilon}[\theta\lambda]_{\lambda_1}^{\lambda_2}-\frac{1}{4G}\int_{\mathcal{C}}\bm{\epsilon}\int_{\lambda_1}^{\lambda_2}\lambda\frac{\mathrm{d}\theta}{\mathrm{d}\lambda}\mathrm{d}\lambda,
\end{align}
where the evolution of the expansion satisfies the null Raychaudhuri equation,
\begin{align}
\frac{\mathrm{d}\theta}{\mathrm{d}\lambda}=-\frac{1}{2}\theta^2-\sigma_{ab}\sigma^{ab}+\omega_{ab}\omega^{ab}-R_{ab}k^ak^b.
\end{align}
Since $k^a$ is orthogonal to the hypersurface, the
rotation tensor $\omega_{ab}$ vanishes. Meanwhile, both $\theta$ and $\sigma$ are of first order in perturbation~\cite{Mishra:2017sqs,Visser:2024pwz}, hence quadratic terms like $\theta^2$ and $\sigma_{ab}\sigma^{ab}$ also vanish in deriving the physical-process first law. After simplification, the evolution of the expansion can be described in terms of the energy-momentum tensor $T_{ab}$,
\begin{align}
\frac{\mathrm{d}\theta}{\mathrm{d}\lambda}=-R_{ab}k^ak^b=-8\pi G T_{ab}k^ak^b.
\end{align}
Substituting the above equation into Eq.~\eqref{eq:IBP}, and considering the dynamical BH entropy to first order in perturbation given in Ref.~\cite{Hollands:2024vbe},
\begin{align}
S_\mathrm{dyn}[\mathcal{C}]=S_\mathrm{BH}[\mathcal{C}]-\frac{1}{4G}\int_{\mathcal{C}}\lambda\theta\bm{\epsilon},
\end{align}
the physical-process first law is  thus obtained as~\cite{Mishra:2017sqs,Hollands:2024vbe},
\begin{align}\label{eq:PPFL0}
\Delta S_\mathrm{dyn}=\frac{2\pi}{\kappa}\int_{\mathcal{C}}\bm{\epsilon}\int_{\lambda_1}^{\lambda_2} T_{ab}\xi^ak^b\mathrm{d}\lambda=\frac{2\pi}{\kappa}\Delta E,
\end{align}
where $\xi^a=\lambda\kappa k^a$ is the killing vector proportional to the horizon generator $k^a$, and $\kappa$ represents the separation to the affine parameter defined as $\xi^b\nabla_b\xi^a\overset{\mathcal{H}}{=}\kappa\xi^a$. It is worth noting that the more general proof given in Ref.~\cite{Hollands:2024vbe} shows that the physical-process first law does not rely on a theory of gravity with a particular dimension, and it applies to an arbitrary classical diffeomorphism-covariant Lagrangian theory of gravity. In what follows, this physical-process first law $\dot{S}\equiv\dot{S}_\mathrm{dyn}=8\pi GM_\mathrm{BH}\dot{M}_\mathrm{BH}$ is only used to relate the energy flow with the entropy flow, on which the information transmission bound acts.

\textit{\textbf{An evaporating BH.}---}
To begin with, Pendry proposed in 1983, $\dot{I}^2\leq\pi|\dot{E}|/3\hbar\ln^22$, between the information (entropy) flow $\dot{I}\equiv\dot{S}/\ln2$ and the energy flow $\dot{E}$, without explicit mention of the Heisenberg time/energy uncertainty relation and independent of statistics and dispersion relationships in use for the channel particles. Naively applying this information-energy flow inequality (Pendry's bound~\cite{Pendry1983} hereafter),
\begin{align}\label{eq:PendryBound}
\dot S^2\leq\frac{\pi}{3\hbar}|\dot E|,
\end{align}
to the early stage of BH evaporation after identifying the system energy $E$ as the BH mass $M_\mathrm{BH}$ and the system entropy $S$ as the dynamical BH entropy with a decreasing rate, $\dot{S}\equiv\dot{S}_\mathrm{dyn}=8\pi GM_\mathrm{BH}\dot{M}_\mathrm{BH}$, given by the physical-process first law~\eqref{eq:PPFL}, one can parametrically reproduce the well-known inverse-squared law of BH mass evaporation rate~\cite{Page:1993wv,Page:2013dx} as an upper bound (upto a numeric prefactor and the greybody prefactor),
\begin{align}\label{eq:dMdt1}
\left|\frac{\mathrm{d}M_\mathrm{BH}/m_\mathrm{Pl}}{\mathrm{d}t/t_\mathrm{Pl}}\right|\leq\frac{1}{192\pi}\frac{m_\mathrm{Pl}^2}{M_\mathrm{BH}^2},
\end{align}
without explicit reference to the Stefan-Boltzmann law. 
Note that the missing greybody factor should emerge from a model-dependent, case-sensitive generalization of Pendry's bound, which was originally derived for noiseless, broadband, single-channel transmission with bosons or fermions. This is reserved for future work.

From the physical-process first law~\eqref{eq:PPFL} and mass evaporation rate~\eqref{eq:dMdt1}, the dynamical entropy decreases as
\begin{align}\label{eq:dSdt1}
\left|\frac{\mathrm{d}S_\mathrm{dyn}}{\mathrm{d}t/t_\mathrm{Pl}}\right|\leq\frac{1}{24}\frac{m_\mathrm{Pl}}{M_\mathrm{BH}}.
\end{align}
It is easy to check both original~\cite{Page:1993wv,Page:2013dx} and recent~\cite{Penington:2019kki,Almheiri:2019qdq} arguments (see also Refs.~\cite{Nian:2019buz,Nian:2023xmr}) for the Page curve of von Neumann entropy, $S_\mathrm{vN}\approx S_\mathrm{rad}\Theta(t_\mathrm{Page}-t)+S_\mathrm{BH}\Theta(t-t_\mathrm{Page})$, interpolating as the minimum of the Hawking radiation entropy $S_\mathrm{rad}=4\pi GM_i^2\beta[1-(1-t/t_\mathrm{age})^{2/3}]$ and BH thermal entropy $S_\mathrm{BH}=4\pi GM_i^2(1-t/t_\mathrm{age})^{2/3}$ before and after the Page time $t_\mathrm{Page}=t_\mathrm{age}\{1-[\beta/(1+\beta)]^{3/2}\}$, all obeys the same inverse-mass law of entropy changing rate~\cite{Page:1993wv,Page:2013dx} (see Refs.~\cite{Guo:2023gfa,Gan:2022jay,Hashimoto:2020cas} for 4D examples), 
\begin{align}\label{eq:dSdt0}
\left|\frac{\mathrm{d}{S}_\mathrm{vN}}{\mathrm{d}t/t_\mathrm{Pl}}\right|\approx8\pi\alpha\frac{m_\mathrm{Pl}}{M_\mathrm{BH}}\left[\beta\Theta(t_\mathrm{Page}-t)+\Theta(t-t_\mathrm{Page})\right],
\end{align}
where $\beta\equiv|\dot{S}_\mathrm{rad}/\dot{S}_\mathrm{BH}|$~\footnote{Strictly speaking, Pendry's bound should apply directly to the channel entropy $S_\mathrm{rad}$ rather than the horizon entropy $S_\mathrm{dyn}$, which differs by a factor $\beta$ of order 1.5 in standard estimates, and hence it can be neglected in the following formulations.}, $\alpha$ is the greybody factor in the traditional inverse-square law of BH mass evaporation rate $\dot{M}_\mathrm{BH}=-\alpha(M_\mathrm{BH})/(GM_\mathrm{BH})^2$, and $t_\mathrm{age}\approx(1/3\alpha(M_i))(M_i/m_\mathrm{Pl})^3t_\mathrm{Pl}$ is an estimate for the lifetime of a BH with an initial mass $M_i$.

However, both our upper bound~\eqref{eq:dSdt1} and the traditional estimate~\eqref {eq:dSdt0} would diverge at the very late stage of BH evaporation, when $M_\mathrm{BH}\to0$. Usually, such a divergence is expected to be rescued by some unknown quantum gravity effects. Nevertheless, as pointed out in Ref.~\cite{Wang:2023wsm}, requiring the Bremermann-Bekenstein bound at this very late stage could enforce the BH mass evaporation rate to be bounded by a constant value before quantum gravity could come into play, if the total number of degrees of freedom of all possible Hawking radiation particles can be enormously large. In simple terms, combing the Bremermann-Bekenstein bound $|\dot{S}|\leq\pi E/\hbar$ with the physical-process first law $\dot{S}\equiv\dot{S}_\mathrm{dyn}=8\pi GM_\mathrm{BH}\dot{M}_\mathrm{BH}$ and system energy $E=M_\mathrm{BH}$ leads to a constant-mass-evaporation-rate as an upper bound,
\begin{align}\label{eq:dMdt2}
\left|\frac{\mathrm{d}M_\mathrm{BH}/m_\mathrm{Pl}}{\mathrm{d}t/t_\mathrm{Pl}}\right|\leq\frac{1}{8},
\end{align}
and a linear rate in the BH mass as an upper bound for the very late-time decrease of dynamical BH entropy, 
\begin{align}\label{eq:dSdt2}
\left|\frac{\mathrm{d}S_\mathrm{dyn}}{\mathrm{d}t/t_\mathrm{Pl}}\right|\leq\pi\frac{ M_\mathrm{BH}}{m_\mathrm{Pl}}.
\end{align}

\begin{figure}
\centering
\includegraphics[width=0.49\textwidth]{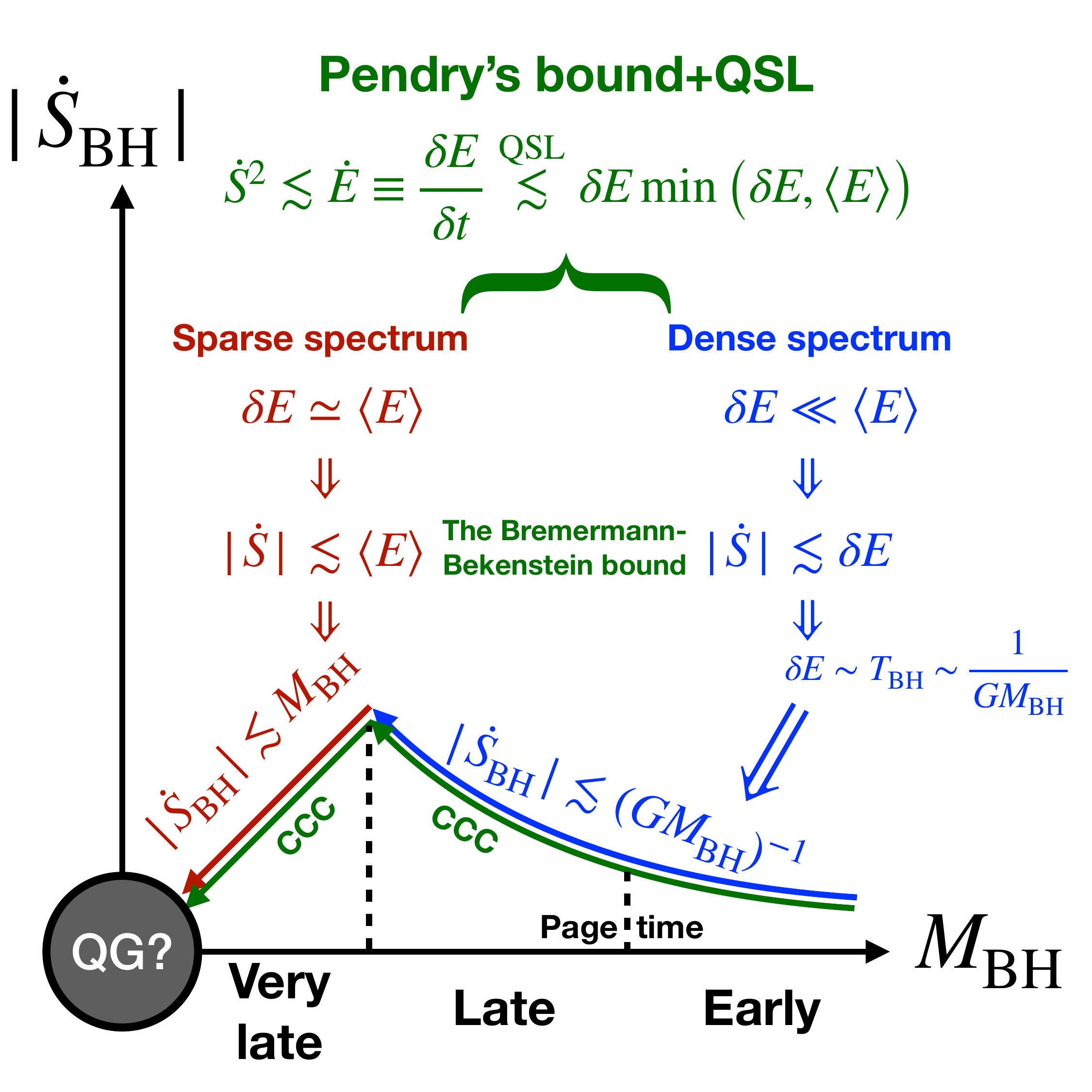}
\caption{The time derivative of black-hole (BH) entropy (green) is conjectured to saturate Pendry's bound limited by the quantum-speed limit (QSL), consistent with the cosmic censorship conjecture (CCC), consisting of both the instantaneous (blue) and average (red) versions of the Bremermann-Bekenstein bounds at early/late or very late stages with dense or sparse energy spectrum, respectively.}
\label{fig:dPage}
\end{figure}

One may question the use of system energy $E=M_\mathrm{BH}$ on the right-hand side of the Bremermann-Bekenstein bound $|\dot{S}|\leq \pi E/\hbar$, as $E$ should be the message energy. In fact, this form is only true at the very late stage of evaporation when the energy spectrum is rather sparse. In general, Pendry's bound, when limited by the QSL, 
\begin{align}
\dot{S}^2\leq\frac{\pi}{3}\frac{\delta E}{\delta t}\leq\frac23\delta E\min(\delta E,\langle E\rangle),
\end{align}
always implies the instantaneous version of the Bremermann-Bekenstein bound with the message energy $\delta E$ on the right-hand side, $|\dot{S}|\leq\sqrt{2/3}\delta E$. This is true for both early and not-too-late stages of Hawking evaporation when the energy spectrum is still very dense so that the Hawking radiation particle energy ($\delta E$) is much smaller than the BH mass (average energy $\langle E\rangle$).  In this case, the emitted message energy is roughly the Hawking temperature, $\delta E\sim T_H\sim1/(GM_\mathrm{BH})$, reproducing the inverse-mass law of entropy flow, $|\dot{S}_\mathrm{BH}|\lesssim\delta E\simeq1/(GM_\mathrm{BH})$. However, with further decrease in BH entropy, the number of microscopic states of the quantum BH also decreases, and the energy spectrum becomes sparse. Until this very late stage of BH evaporation, when the energy spread between neighboring states is of the same order as the average energy, $\delta E\sim\langle E\rangle$, Pendry's bound limited by the QSL roughly reproduces the average version of the Bremermann-Bekenstein bound with the average energy on the right-hand side, $|\dot{S}|\lesssim\langle E\rangle$. Therefore, the entropy flow should always follow Pendry's bound limited by the QSL throughout the evaporation history (see Fig.~\ref{fig:dPage}). 

Note that in this unified picture, the physical-process first law $\dot{S}\equiv\dot{S}_\mathrm{dyn}=8\pi GM_\mathrm{BH}\dot{M}_\mathrm{BH}$ is not explicitly used when deriving the entropy flow bounds $|\dot{S}_\mathrm{BH}|\lesssim1/(GM_\mathrm{BH})$ for the dense spectrum and $|\dot{S}_\mathrm{BH}|\lesssim M_\mathrm{BH}$ for the sparse spectrum, but is explicitly used to convert the entropy flow bounds~\eqref{eq:dSdt1} and~\eqref{eq:dSdt2} to the energy flow bounds~\eqref{eq:dMdt1} and~\eqref{eq:dMdt2}. Therefore, in both cases that assume saturation of the mass evaporation laws~\eqref{eq:dMdt1} and~\eqref{eq:dMdt2}, the Pendry's bound~\eqref{eq:dSdt1} and Bremermann-Bekenstein bound~\eqref{eq:dSdt2} would all be consistent with the physical-process first law that is compatible with the Penrose inequality~\cite{Geroch:1973,Jang:1977,Huisken:2001,Hubert:2001,Bray:2003ns,Bray:2007opu} for the apparent horizon $A_\mathrm{app}=4GS_\mathrm{dyn}$~\cite{Hollands:2024vbe} as one can explicitly check,
\begin{align}
M_\mathrm{BH}\geq\sqrt{\frac{A_\mathrm{app}}{16\pi G^2}}.
\end{align}
Note that the Penrose inequality here only serves as a consistency property of the physical-process first law, and by no means either proves the general Penrose inequality or determines whether a singularity is globally censored. A genuinely dynamical claim would need to invoke the known critical rate separating naked from covered singularities in self-similar Vaidya spacetimes~\cite{Barve:2000wm,Mkenyeleye:2014dwa}.

The slowing-down evaporation rate~\eqref{eq:dSdt2} compared to earlier would-be divergent rate~\eqref{eq:dSdt0} suggests a transition period around $M_\mathrm{BH}=\sqrt{8\alpha} m_\mathrm{Pl}$ if the slowing-down evaporation rate~\eqref{eq:dSdt2} or~\eqref{eq:dMdt2} is conjectured to be saturated. Recall that in Ref.~\cite{Ukwatta:2015iba},
\begin{align}
\alpha=\frac{27}{2\pi(8\pi)^4}\sum_ig_{\mathrm{dof},i}\left(\int_0^\infty\frac{\gamma_{s_i}(x)x^3\mathrm{d}x}{e^x-(-1)^{2s_i}}\equiv\Phi_i\sim\mathcal{O}(1)\right),
\end{align}
this conjectured accelerating-to-constant evaporation period can occur even before quantum gravity effects could come into play if $\alpha\sim10^{-5}\sum_ig_{\mathrm{dof},i}\Phi_i\sim\mathcal{O}(10^{-4}g_\mathrm{dof})\gg1/8$, that is, the total number of degrees of freedom of all possible Hawking radiation particles to be emitted is sufficiently large, $g_\mathrm{dof}>10^4$, as expected from some scenarios beyond the standard model of particle physics, such as supersymmetry, moduli, and axion-like particles~\cite{Baker:2022rkn}, $N$-naturalness~\cite{Arkani-Hamed:2016rle}, and large $N$ species solutions to the hierarchy problem~\cite{Dvali:2007hz,Dvali:2007wp,Calmet:2008tn,Dvali:2009ne}.

\textit{\textbf{An expanding dS.}---} 
Similar to an evaporating BH, we can apply the Bremermann-Bekenstein bound to a dS space to see what it would lead to. Recall that the entanglement entropy of a free massless scalar field at its nondegenerate ground state in $D=1+3$ dimensional flat space scales as
\begin{align}
S_\mathrm{flat}=d_2\frac{R^2}{\epsilon^2}+d_1\ln\frac{R}{\epsilon}+d_0+\cdots
\end{align}
in terms of an ultraviolet (UV) cutoff length $\epsilon$, whose continuum limit would think less of those terms $(\epsilon/R)^{2i}$ at higher orders. The leading-order area-law term found in the seminal work~\cite{Srednicki:1993im} by Srednicki bears a striking similarity to the BH entropy if the UV cutoff scale is set at the reduced Planck scale $M_\mathrm{Pl}^2\equiv1/8\pi G$. The first sub-leading correction is a logarithmic term in $R/\epsilon$ with a universal coefficient $d_1=-1/90$~\cite{Solodukhin:2008dh,Casini:2009sr} from holographic arguments that was later confirmed numerically~\cite{Lohmayer:2009sq}. 

When considered in the dS space, the entanglement entropy for a subsystem of physical size $R_p=aR$ admits an extra dependence on the Hubble horizon scale $H=-1/(a\tau)$, where $a$ is the scale factor, $R$ is the comoving scale, and $\tau$ is the conformal time. At the beginning of inflation with $\tau\to-\infty$, all systems are of sub-horizon size $R_p\ll1/H$ so that the entanglement entropy should reproduce the flat-space one. For a subsystem exactly of horizon size $R_p=1/H$, it was expected from the usual thermal density matrix in the static patch~\cite{Maldacena:2012xp} and the associated entropy can be regarded as a (UV divergent) $\mathcal{O}(G_N^0)$ correction to the gravitational entropy of dS space~\cite{Callan:1994py}. For a super-horizon-sized spherical surface, $R_p>1/H$, the entanglement entropy of a free massive scalar field was suggested to be of the form~\cite{Maldacena:2012xp}
\begin{align}
S_\mathrm{dS}=c_1\frac{R_p^2}{\epsilon_p^2}&+\ln(H\epsilon_p)(c_2+c_3m^2R_p^2+c_4H^2R_p^2)\nonumber\\
&+c_5H^2R_p^2-c_6\ln(HR_p)+\mathrm{const.},
\end{align}
where a consistency check has been done for a conformally coupled scalar field to reproduce the flat-space term $d_1\ln(R/\epsilon)$ with $c_2=c_6=1/90$ since the dS space is conformally flat. In particular, in $D=1+3$ dimensional dS space, the interesting piece of the entanglement entropy is proportional to the number of e-foldings that elapsed~\cite{Maldacena:2012xp} since the spherical region was once inside the horizon. We will see this result once again shortly below.

For a sub-horizon size entangling surface, $R_p< 1/H$, a more general form was suggested as~\cite{Boutivas:2024sat,Boutivas:2024lts}
\begin{align}
S_\mathrm{dS}&=\frac{R_p^3}{\epsilon_p^3}f_3(HR_p)+\frac{R_p^2}{\epsilon_p^2}f_2(HR_p)+\frac{R_p}{\epsilon_p}f_1(HR_p)\nonumber\\
&+f_0(HR_p)+g(HR_p)\ln\frac{R_p}{\epsilon_p}+h(HR_p)\ln\frac{L_p}{\epsilon_p},
\end{align}
where a brute-force numerical method has been adopted to confirm (i) the absence of volume term, $f_3(HR_p)=0$; (ii) no new UV divergences, $f_2(HR_p)=d_2$, $f_1(HR_p)=0$; (iii) the dS expansion effects encoded in $f_0(HR_p)=d_0+a_2^{(2)}H^2R_p^2$ and $g(HR_p)=-1/90+a_2^{(2)'}H^2R_p^2$ with $a_2^{(2)}=-0.142650$ and $a_2^{(2)'}=0.000025$; and (iv) a new interesting dependence on the infrared (IR) cutoff  $L_p$ of the overall system with a universal coefficient~\cite{Boutivas:2024sat,Boutivas:2024lts},
\begin{align}
h(HR_p)=\frac13H^2R_p^2,
\end{align}
which has also been confirmed analytically as the leading correction to the flat-space result. Note that the absence of explicit $\ln(H\epsilon_p)$ term could be re-introduced by splitting the $\ln(R_p/\epsilon_p)$ term into $\ln(HR_p)-\ln(H\epsilon_p)$ and absorbing $g(HR_p)\ln(HR_p)$ into $f_0(HR_p)$. However, numerical estimates found no such logarithmic term in $f_0(HR_p)$ since the sub-horizon form of entanglement entropy is expanded at early time around flat space with $HR_p<1$, which precludes the appearance of such terms in the first place. A late-time expansion of the entanglement entropy might as well combine such terms back to $\ln(H\epsilon_p)$ just as the super-horizon form. Remarkably, the presence of such an IR-dependent term $(1/3)H^2R_p^2\ln(L_p/\epsilon_p)$ goes beyond the region of validity of perturbation theory at early time around flat space, as it also appears during the exact evaluation for the entanglement entropy~\cite{Boutivas:2024sat,Boutivas:2024lts}.

The IR cutoff length $L_p$ of the overall system can be roughly chosen as the wavelength of the first mode that exited the horizon just right after inflation started at $a_i$, which would extend far beyond the horizon at a later time $a_f$, that is $L_p=(H^{-1}/a_i)a_f=e^N/H$ with $a_f=a_ie^N$. Then, the IR-dependent term contributes as $(1/3)H^2R_p^2[N-\ln(H\epsilon_p)]$, partially reproducing the aforementioned term proportional to the number of e-foldings that elapsed. Thus, for a pure dS space with a constant $H=\mathrm{d}N/\mathrm{d}t$, the rate of change in $S_\mathrm{dS}$ for the subsystem of sub-horizon size $R_p$ receives contributions only from the IR-dependent term but is independent of specific $L_p$,
\begin{align}
\frac{\mathrm{d}S_\mathrm{dS}}{\mathrm{d}t}=\frac13H^3R_p^2<\frac13H,  \quad\text{for all } R_p<H^{-1}.
\end{align}
Next, to impose the instantaneous version of the Bremermann-Bekenstein bound, one has to figure out the amount of message energy that goes into this subsystem since the entanglement entropy is in fact enhanced due to the squeezed states under the dS expansion, during which super-UV modes of smaller lengths at early time, $\lambda_p<\epsilon_p$, are redshifted at the late time into the effective-field-theory (EFT) regime~\cite{Maldacena:2012xp}, $\epsilon_p<(\lambda_p/a_i)a_f<L_p$. To ensure the validity of the EFT description of inflation, the super-UV mode that just comes into the EFT regime should at least exclude those trans-Planckian modes, $(M_\mathrm{Pl}^{-1}/a_i)a_f\leq\epsilon_p$. Since the new mode that just enters the EFT regime admits exactly the amount of energy $E_p=2\pi\hbar/\epsilon_p$, imposing the instantaneous version of the Bremermann-Bekenstein bound $\mathrm{d}S_\mathrm{dS}/\mathrm{dt}\leq \pi E_p/\hbar$ for all sub-horizon regions would require
\begin{align}\label{eq:dSdtall}
\frac13H\leq\frac{\pi E_p}{\hbar}=\frac{2\pi^2}{\epsilon_p}\leq2\pi^2\frac{M_\mathrm{Pl}}{a_f/a_i}=2\pi^2\frac{M_\mathrm{Pl}}{e^N},
\end{align}
which reduces to a weaker version of the trans-Planckian censorship conjecture~\cite{Bedroya:2019snp,Bedroya:2019tba,Cai:2019dzj} up to a numeric prefactor,
\begin{align}\label{eq:TCC}
e^N\lesssim6\pi^2\frac{M_\mathrm{Pl}}{H}.
\end{align}
Here, the requirement $H/3\leq\pi E_p/\hbar$ in Eq.~\eqref{eq:dSdtall} serves as a sufficient condition to impose the Bremermann-Bekenstein bound for all sub-horizon regions of dS expansion.
On the other hand, if the Bremermann-Bekenstein bound is to be violated, $\mathrm{d}S_\mathrm{dS}/\mathrm{dt}> \pi E_p/\hbar$, and there is at least one trans-Planckian mode destroying the EFT description, $(M_\mathrm{Pl}^{-1}/a_i)a_f>\epsilon_p$, then the trans-Planckian censorship conjecture is also violated,
\begin{align}
2\pi^2\frac{M_\mathrm{Pl}}{e^N}<\frac{2\pi^2}{\epsilon_p}=\frac{\pi E_p}{\hbar}<\frac{\mathrm{d}S_\mathrm{dS}}{\mathrm{d}t}<\frac{H}{3}\Rightarrow e^N>6\pi^2\frac{M_\mathrm{Pl}}{H}.
\end{align}
Therefore, the trans-Planckian censorship conjecture is consistent with the Bremermann-Bekenstein bound in the sense of a valid EFT description of dS expansion.

In analogy to the BH case as a transmitter of information (loss of horizon entropy through Hawking evaporation), we phrase this property of dS expansion as a receiver of information (growth of accessible effective degrees of freedom during inflation). Here, ``receiver'' is not meant to describe dS spacetime as an active physical detector receiving a message from an external sender. Instead, it refers to the effective low-energy description available to a comoving observer in an inflating Hubble patch. Modes that were initially above the UV cutoff are redshifted into the EFT regime. In this limited sense, the EFT Hilbert space gains access to degrees of freedom that were previously outside its domain of validity. Hence, imposing the Bremermann-Bekenstein bound on all sub-horizon regions of dS expansion is to bound the rate at which new effective degrees of freedom enter the low-energy description in such a way that is consistent with the trans-Planckian censorship conjecture up to a numeric prefactor.

\textit{\textbf{Conclusions and discussions.}---}
Instead of proving respectively that an evaporating Schwarzschild BH and an expanding dS space are the fastest transmitter and receiver of information in nature, we discuss in this Letter only their corresponding immediate consequences. For the evaporating BH case, imposing the Pendry's bound limited by the QSL on the decreasing rate of BH entropy tends to slow down the upper bounds on BH mass evaporation from the usual accelerating rate to a constant rate, which is consistent with the Penrose inequality from the cosmic censorship conjecture. For the expanding dS case, imposing the Bremermann-Bekenstein bound on all sub-horizon regions of dS space with an EFT description tends to bound the rate at which new EFT-accessible degrees of freedom can be absorbed into an EFT description of inflation, which is consistent with the trans-Planckian censorship conjecture up to some numeric factor.


The compatibility between the information-flow bounds and the Penrose inequality provides a consistency check between the information-theoretic description and cosmic censorship. Our argument does not constitute an independent derivation of cosmic censorship.
Furthermore, our explicit use of dynamical BH entropy from the apparent horizon could lead to a failure without time symmetry~\cite{Ben-Dov:2004lmn}. This might as well suggest the dynamical BH entropy may be defined by the minimum area enclosing the apparent horizon beyond the leading-order perturbation analysis.

Although we have limited our discussions within the semi-classical regimes $M_\mathrm{BH}\gg m_\mathrm{Pl}$, there is no guarantee that all semi-classical regimes are quasi-stationary so that the physical-process first law can always be applied as a first-order perturbation. However, recent studies~\cite{Ashtekar:2025qqa,Ashtekar:2026jdz} extend the first law to encompass BHs that can be arbitrarily far from equilibrium. It now refers to finite changes that occur due to physical processes, which reduce to the physical-process first law locally.

The trans-Planckian censorship conjecture is originally imposed by forbidding the trans-Planckian modes from being stretched out of the Hubble horizon to be classically probed in the subsequent cosmic history; that is, all modes that exit the Hubble horizon at the end of inflation should have their physical length larger than the Planck length at the beginning of inflation, $a_i(a_fH)^{-1}>M_\mathrm{Pl}^{-1}$, so that $e^N<M_\mathrm{Pl}/H$. However, our derivations only forbid the stretched trans-Planckian modes from entering into the EFT description. As the UV cutoff scale is usually larger than the Hubble scale, our approach to the trans-Planckian censorship conjecture from the Bremermann-Bekenstein bound is much weaker (as the bound is slightly looser due to an extra $6\pi^2$ prefactor) and thus more general than the original argument. Although our use of the dS entanglement entropy neglects the slow time evolution of the Hubble scale, a reconsideration should simply recover the refined trans-Planckian censorship conjecture~\cite{Cai:2019dzj}.

\begin{acknowledgments}
We thank Yu-Sen An, Liming Cao, Jun Nian, Li Li, Shan-Ming Ruan, Dong-Gang Wang, Houwen Wu, Haitang Yang, Run-Qiu Yang, and Hongbao Zhang for helpful discussions.
This work is supported by the National Key Research and Development Program of China Grants No. 2021YFC2203004, No. 2021YFA0718304, and No. 2020YFC2201501, the National Natural Science Foundation of China Grants No. 12422502, No. 12547110, No. 12588101, No. 12235019, and No. 12447101.
\end{acknowledgments}


\bibliography{ref}

\end{document}